\documentstyle[12pt,aaspp4,psfig]{article}
\newcommand{\etal}{{\it et al.~}}
\newcommand{\Halpha}{H$\alpha$}

\begin{document}

\title{THE GLOBAL STAR FORMATION RATE FROM THE 1.4~GHz LUMINOSITY FUNCTION}
\author{L. E. CRAM} \affil{School of Physics, University of Sydney NSW
  2006, Australia; L.Cram@physics.usyd.edu.au}

\begin{abstract}

  The decimetric luminosity of many galaxies appears to be dominated by
  synchrotron emission excited by supernova explosions. Simple models
  suggest that the luminosity is directly proportional to the rate of
  supernova explosions of massive stars averaged over the past $\approx 3
  \times 10^7$ yr.  The proportionality may be used together with models of
  the evolving 1.4~GHz luminosity function to estimate the global star
  formation rate density in the era $z \lesssim 1$. The local value is
  estimated to be 0.026 M$_\odot$ yr$^{-1}$ Mpc$^{-3}$, some 50\% larger
  than the value inferred from the \Halpha~ luminosity density. The value
  at $z \approx 1$ is found to be 0.30 M$_\odot$ yr$^{-1}$ Mpc$^{-3}$. The
  10-fold increase in star formation rate density is consistent with the
  increase inferred from  mm-wave, far-infrared, ultra-violet and \Halpha~
  observations.

\end{abstract}


\section{INTRODUCTION}

Two different mechanisms are believed to energize the relativistic
electrons responsible for the decimetric synchrotron emission from
galaxies.  In classical radio galaxies and related objects, the energy
source is ultimately a non-stellar central engine, perhaps a massive black
hole. In normal disk galaxies and related objects, the energy source is
ultimately stellar, perhaps the totality of supernova explosions over the
past $\approx 3 \times 10^7$ yr. Although the relative importance of the
two mechanisms in some galaxies is debated, it is usually not difficult to
determine the dominant source in any galaxy from observations of its radio
morphology, and the brightness, morphology and spectrum of any optical
counterpart.

For galaxies in which the radio emission is energized by stars (hereinafter
star-forming galaxies) it should be possible to infer the underlying rate
of star formation from the observed radio luminosity. Because the radio
luminosity function of star-forming galaxies is well determined locally,
and measurable out to $z \approx 1$, and because radio emission usually
does not suffer absorption, the derivation of star formation rates from
radio luminosities offers an important addition to other methods used to
determine global star formation rates (e.g., Gallego \etal
\markcite{Gal95}1995; Rowan-Robinson \etal \markcite{Row97}1997; Pettini
\etal \markcite{Pet97}1997; Tresse \& Maddox \markcite{Tre98}1998; Treyer
\etal \markcite{Tre98}1998; Blain \etal \markcite{Bla98}1998).  On the
other hand, the {\it quantitative} link between 1.4~GHz luminosity and the
rate of star formation is rather tenuous, and further work is
clearly needed to strengthen the astrophysical basis of the method.

This note derives the star formation rate density for $z \lesssim 1$ from
the observed radio luminosity function, and compares the result with other
determinations of this important quantity. A Hubble constant $H_0 =
50$~km~s$^{-1}$~Mpc$^{-1}$ has been adopted, and previously published
quantities dependent on $H_0$ have been scaled if necessary.

\section{THE STAR FORMATION RATE AT $z \approx 0$} 

Condon (\markcite{Con84a}1984a, \markcite{Con84b}1984b,
\markcite{Con88}1988, \markcite{Con89}1989) has determined the local radio
luminosity function for the combined population of spiral and irregular
galaxies in the form
\begin{equation}
  \phi_m (L) = \log_e (10^{0.4}) \left ( \frac{L}{L_0} \right )^{3/2}
  10^{\left ( 3.15 - \left [ \frac{\log_{10} (L) - 22.0}{1.9} \right ]^2
    \right )}.
\end{equation}
Here, $L$ is the luminosity (W Hz$^{-1}$) at 1.4~GHz, $ \phi_m = \log_e (m)
L \phi (L) $ is the luminosity function per unit interval in $\log_m (L)$
and per unit volume (Mpc$^{-3}$), and $L_0 = 10^{23.54}$ is the factor
converting between SI units and astronomical units. The corresponding
1.4~GHz luminosity density, ${\cal L}_{1.4}$, is given by

\[
{{\cal L}_{1.4}} = \int_o^\infty L \phi (L) {\rm d}L = 1.8 \times 10^{19}~
{\rm W~Hz}^{-1}~{\rm Mpc}^{-3}.
\]
The uncertainty in this estimate (Condon \markcite{Con89}1989, Table 3) is
a factor of $10^{\pm 0.2}$.

Condon and Yin \markcite{Con90}(1990) and Condon \markcite{Con92b}(1992)
have determined a relationship between the 1.4~GHz luminosity of a galaxy
and its current star formation rate (SFR) which can be expressed in the
form
\begin{equation}
  SFR_{1.4} = \left( \frac{L_{1.4}}{2.35 \times 10^{22}~{\rm W Hz}^{-1}}
  \right)~~~{\rm M}_{\odot}~{\rm yr}^{-1}.
\end{equation}
The conversion factor in Equation (2) corresponds to an adopted initial
mass function $\Psi(M) \sim M^{-2.35}$ with $0.1 < M < 100$ M$_{\odot}$.
Equations (1) and (2) imply that the local value of the global star
formation rate density is $0.026$ M$_\odot$ yr$^{-1}$ Mpc$^{-3}$, with an
uncertainty of the order of a factor of $10^{\pm 0.2}$.  The most luminous
galaxy encountered in constructing the local luminosity function of the
star-forming population has a 1.4 GHz luminosity of $\approx 10^{24.4}$ W
Hz$^{-1}$ (Condon \markcite{Con89}1989, Table 3); its corresponding
inferred star formation rate is just over 100 M$_\odot$ yr$^{-1}$.

\section{THE STAR FORMATION RATE FOR $z \lesssim 1$}

Sub-mJy radio surveys not suffering from confusion were first made by
aperture synthesis telescopes in the early 1980s.  The shape of the 1.4 GHz
source-count distribution revealed by these studies provided compelling
evidence for significant evolution of the star-forming galaxy population.
As reviewed by Condon \markcite{Con88}(1988), the source count distribution
normalised to the Euclidean slope [i.e., the curve $N(S) S^{-2.5}$]
flattens for $S_{1.4} \lesssim 5 $ mJy.  This flattening {\it cannot} be
interpreted in terms of the population of AGN-powered galaxies which
dominate the source counts above 10 mJy, nor can it arise from a
non-evolving population of star-forming galaxies having the local
luminosity function presented above (Danese \etal \markcite{Dan87}1987).
The faint counts can be reconciled with the local luminosity function of
star-forming galaxies by allowing significant evolution of the star-forming
population in the interval $z \lesssim 1$. Danese \etal
\markcite{Dan87}(1987; see also Rowan-Robinson \etal \markcite{Row93}1993;
Hopkins \etal \markcite{Hop98}1998) have shown that consistency between the
local luminosity function and the faint source counts for the star-forming
population can be obtained by a pure luminosity evolution model satisfying
$P_{1.4}(z) \propto (1 + z)^3$.

In studies of the entire radio source population, Condon (1984a, 1984b,
1988, 1989) has established consistency between the evolving luminosity
functions of both star-forming and AGN galaxies and the radio-frequency
source counts over six orders of magnitude of flux density.  His model uses
a density evolution factor $g(z)$ and a luminosity evolution factor $f(z)$
in the form
\[
\phi_m (L, z) = g(z) \phi_m [L/f(z), z=0].
\]
This is the `co-moving' luminosity function; Condon adopted a cosmological
model with $\Omega_0=1$ and $h = 0.5$. Values of $f(z)$ and $g(z)$ taken
from Condon (\markcite{Con84b}1984b, Table 1), combined with Equations (1)
and (2), lead to the $z-$dependence of the global star formation rate
density illustrated in Figure 1. The curve is consistent with that which
would be derived using pure luminosity evolution $P(z) \propto (1+z)^3$.

Although Condon's (1984b) model establishes consistency between the
evolving luminosity function and the source counts, it is not unique.
Nevertheless, source-count models involving significant evolution of the
population of sub-mJy galaxies find growing support as studies of their
optical counterparts show signs of vigorous star-formation. For example,
Kron, Koo \& Windhorst \markcite{Kro85}(1985) found that the optical
counterparts of the sub-mJy population include a blue component, dubbed `B
radio galaxies', with peculiar optical morphologies suggestive of
interactions and mergers.  Benn \etal \markcite{Ben93}(1993) have shown
that many faint radio sources appear to be more-distant counterparts of the
IRAS star-forming galaxy population. This population probably overlaps with
the `B radio galaxies'.  {\it Hubble Space Telescope} images suggest that
the majority of sub-mJy radio sources are starburst or post-starburst
objects, embedded in luminous disk galaxies and triggered by interactions
(Windhorst \etal \markcite{Win95}1995).

\section{COMPARISON WITH OTHER ESTIMATES}

Gallego \etal \markcite{Gal95}(1995) have estimated the global star
formation rate at $z \approx 0$ from their observations of the local
\Halpha~luminosity density.  Their derived star formation rate density is
0.013 M$_\odot$ yr$^{-1}$ Mpc$^{-3}$, one-half the value derived above.
While a factor of two uncertainty in the global star formation density
derived by such different methods is not disturbingly large, it is of
interest to explore the origin of the difference. Accordingly, Figure 2
compares the 1.4~GHz and \Halpha~luminosity functions and the corresponding
distributions of the inferred star formation rate densities, both expressed
on a common $\log (SFR)$ scale to facilitate comparison.

Over the range where the \Halpha~luminosity function is well determined ($4
\times 10^{33} < L_{H\alpha} < 4 \times 10^{35}$ W or $0.4 < SFR_{\alpha} <
40$ M$_{\odot}$ yr$^{-1}$) the two luminosity functions cross, with the
\Halpha~distribution rising somewhat more steeply to lower luminosities and
dropping slightly below the 1.4~GHz function at the highest observed point.
At faint luminosities the extrapolation of the \Halpha~luminosity function
rises above the 1.4~GHz function, but the difference never exceeds a factor
of 5 even at the faintest luminosities. The star formation rate density
inferred from \Halpha~is approximately twice the density inferred from
1.4~GHz observations, for values of SFR below about $40$ M$_{\odot}$
yr$^{-1}$.

The excess of the 1.4~GHz value of SFR relative to the \Halpha~value
therefore arises at high values of SFR. This is clear from Figure 2, which
shows that the \Halpha~ luminosity function is $\approx 10^{-6}$ dex$^{-1}$
Mpc$^{-1}$ at $SFR_{\alpha} = 100$ M$_{\odot}$ yr$^{-1}$ while at the same
$SFR_{1.4}$, the 1.4~GHz luminosity function is a factor of 25 greater.
Luminosity weighting amplifies this difference in the plot of the star
formation rate density. As a result, the upper quartile of star formation
rate densities lies above 25 M$_\odot$ yr$^{-1}$ according to the 1.4~GHz
data, and above only 15 M$_\odot$ yr$^{-1}$ for the \Halpha~data. Even more
extreme is the high star formation rate tail which extends to values $>
100$ M$_\odot$ yr$^{-1}$ in the 1.4~GHz data, but is absent in \Halpha~(see
also Cram \etal 1998; objects in this category include Arp~220 and
NGC~6240).  The excess SFR density in the 1.4~GHz data at high values of
SFR more than outweighs the excess \Halpha~SFR density at low values of
SFR, and accounts for the somewhat higher global rate deduced from the 1.4~GHz
data.

Other estimates of the global star formation rate at low redshift ($z
\lesssim 0.3$) have been made by Treyer \etal \markcite{Try98}(1998) and
Tresse \& Maddox \markcite{Trs98}(1998).  The former, based on
balloon-borne observations at 200 nm, yields 0.006 M$_{\odot}$ yr$^{-1}$
(adjusted to $h=0.5$) at $z \approx 0.15$, and the latter, based on
\Halpha~spectra from the $I$-selected Canada-France Redshift Survey, yields
0.020 M$_{\odot}$ yr$^{-1}$ at $z \approx 0.2$.  The \Halpha~ value is
compatible with that of Gallego \etal (1995), while the far-UV value is
smaller, perhaps significantly so. Both are smaller than the 1.4~GHz value.

There have been several determinations of the evolution of the SFR density
beyond $z \approx 0.25$ (e.g., Lilly \etal \markcite{Lil96}1996; Madau
\etal \markcite{Mad96}1996; Connolly \etal \markcite{Con97}1997;
Rowan-Robinson \etal \markcite{Row97}1997). All agree that there is a
significant rise between $z = 0$ and $z = 1$, of a factor of order 10. The
1.4~GHz results are consistent with this trend.

It should be noted that dust extinction leads to uncertainty in estimates
of the star formation rate from UV and optical indicators. For example,
Pettini \etal (\markcite{Pet97}1997, their Fig 7) suggest that an upward
revision of a factor of approximately 2 - 6 should be applied to
uncorrected estimates of star formation rates based on ultraviolet
observations. Their revised values lie close to, but below, those estimated
in this paper in the range $0.2 \lesssim z \lesssim 1$. Blain \etal
(\markcite{Bla98}1998, their Figure 8) have estimated the evolution of the
global star formation rate using sub-mm observations which, like radio
observations, should be unaffected by extinction. Their estimates agree, to
within a factor of two over the range $z \lesssim 1$, with those derived here.

\section{DISCUSSION AND SUMMARY}

In addition to the consistency with sub-mm observations noted above, there
is close agreement between the evolution of the SFR density derived from
1.4~GHz data, and from far-infrared observations of the {\it Hubble Deep
  Field} by the Infrared Space Observatory (Rowan-Robinson \etal
\markcite{Row97}1997).  This result could have been anticipated, in view of
the strong correlation between 1.4~GHz and far-infrared luminosities in
local star-forming galaxies, and the viability of theories of the
correlation which relate both luminosities to the number of massive stars
in the galaxies (e.g. Condon 1992). Estimates of the SFR density inferred
from far-UV or \Halpha~data (uncorrected for extinction) appear to be
systematically lower, by a factor of order $2-4$.  The discrepancy could be
a consequence of a moderate amount of optical/far UV extinction, an
interpretation that finds some support in studies of nearby galaxies which
have revealed evidence of widespread extinction, including high attenuation
in some objects (e.g., Young, Kleinmann \& Allen \markcite{You88}1988; see
also Pettini \etal \markcite{Pet97}1997). However, it should also be noted
that the ratio of the instantaneous luminosity of various SFR indicators to
the short-term average of the SFR itself is a complex function of time
(e.g., \Halpha~is weighted strongly to the most massive stars' lifetimes,
while 1.4 GHz radiation is weighted to the more abundant stars lying just
above the lower mass limit for supernovae). It is likely that factors other
than extinction will also prove to be relevant in explaining the
discrepancy between various SFR indicators.

Further studies of the extinction in galaxies having high values of
$SFR_{1.4}$ and low values of $SFR_\alpha$ could help resolve this
question.  Of particular interest are galaxies having $SFR_{1.4} \gtrsim
25$ M$\odot$ yr$^{-1}$. In such galaxies, the `shortfall' in
\Halpha~emission is particularly large, and this might lead to obvious
characteristic differences between the morphology in 1.4~GHz and
\Halpha~images.

In Condon's (1984b) model the evolution of the star-forming galaxy
luminosity function is constrained almost exclusively by the fit to sub-mJy
1.4~GHz source counts. This is not a particularly tight constraint, and as
more redshifts become available for sub-mJy star-forming galaxies the
evolution will become better defined. Already, the studies of Benn \etal
\markcite{Ben93}(1993), Rowan-Robinson \etal \markcite{Row93}(1993) and
Hopkins \markcite{Hop98b}(1998) suggest that consistency arises
between the local 1.4~GHz luminosity function of star-forming galaxies, the
sub-mJy source counts and the available $n(z)$ distributions if there is
luminosity evolution according to $(1+z)^3$. This picture is consistent
with results obtained at other wavelengths (e.g. Hacking, Condon \& Houck
\markcite{Hac87}1987). Since the 1.4~GHz data suggest that approximately
50\% of star formation at $z=1$ takes place in star forming galaxies with
$L_{1.4} \gtrsim 10^{24}$ W Hz$^{-1}$, and since such galaxies are not
difficult to detect at radio frequencies ($S_{1.4} \approx 250 \mu$Jy),
optical spectroscopy of complete samples of sub-mJy 1.4~GHz sources could
make an important contribution to our understanding of cosmic star
formation history in the era $z \lesssim 1$.

It is worth noting that studies of the cosmic evolution of {\it powerful}
radio galaxies (i.e., AGN-powered objects with $P_{1.4 {\rm GHz}} >
10^{26}$ W Hz$^{-1}$) suggest that the population undergoes pure luminosity
evolution according to $P_{1.4}(z) \propto (1 + z)^3$ out to $z \approx 2$,
with a luminosity dependent redshift cut-off at higher $z$ (e.g. Dunlop \&
Peacock \markcite{Dun90}1990). As emphasised by Dunlop
(\markcite{Dun98}1998), these studies provide a robust determination of the
evolution of the 1.4~GHz luminosity density, despite remaining
uncertainties in the high-$z$ evolution of the luminosity function itself.
It is a remarkable fact that the luminosity evolution of the AGN-powered
population appears to be quite similar that of the star-forming population
(Danese \etal \markcite{Dan87}1987; Dunlop 1998). Dunlop (1998) advances
the suggestion that the star formation rate and the AGN (black hole)
fueling rate might both be related to the global rate of gravitational
accretion/condensation.

The use of 1.4 GHz radiation to determine star formation rates offers some
advantages over alternative methods. For example, 1.4~GHz radiation does
not usually suffer absorption, and the local 1.4~GHz luminosity function of
star-forming galaxies is well determined over a wide luminosity range.
Compared with mm-wave and far-infrared (FIR) indicators, which share these
advantages over optical and UV indicators, deep 1.4 GHz observations have
superior sensitivity, and offer astrometric precision sufficient to make
direct optical identifications of even the faintest sources.  The main
weakness is the tenuous linkage between 1.4~GHz radiation and the
astrophysics of star formation.  Spatially resolved observations probing
the relationship between decimetric radio emission and star forming
processes in selected galaxies (e.g. Duric \etal \markcite{Dur95}1995;
Marsh \& Helou \markcite{Mar98}1998; Smith \etal \markcite{Smi98}1998) tend
to support the model used above, but further work is clearly needed to
strengthen the astrophysical basis of the method.

\newpage 



\newpage
\begin{center}
\begin{minipage}{15cm}
  \psfig{figure=Fig1.ps,angle=-90,width=15cm}
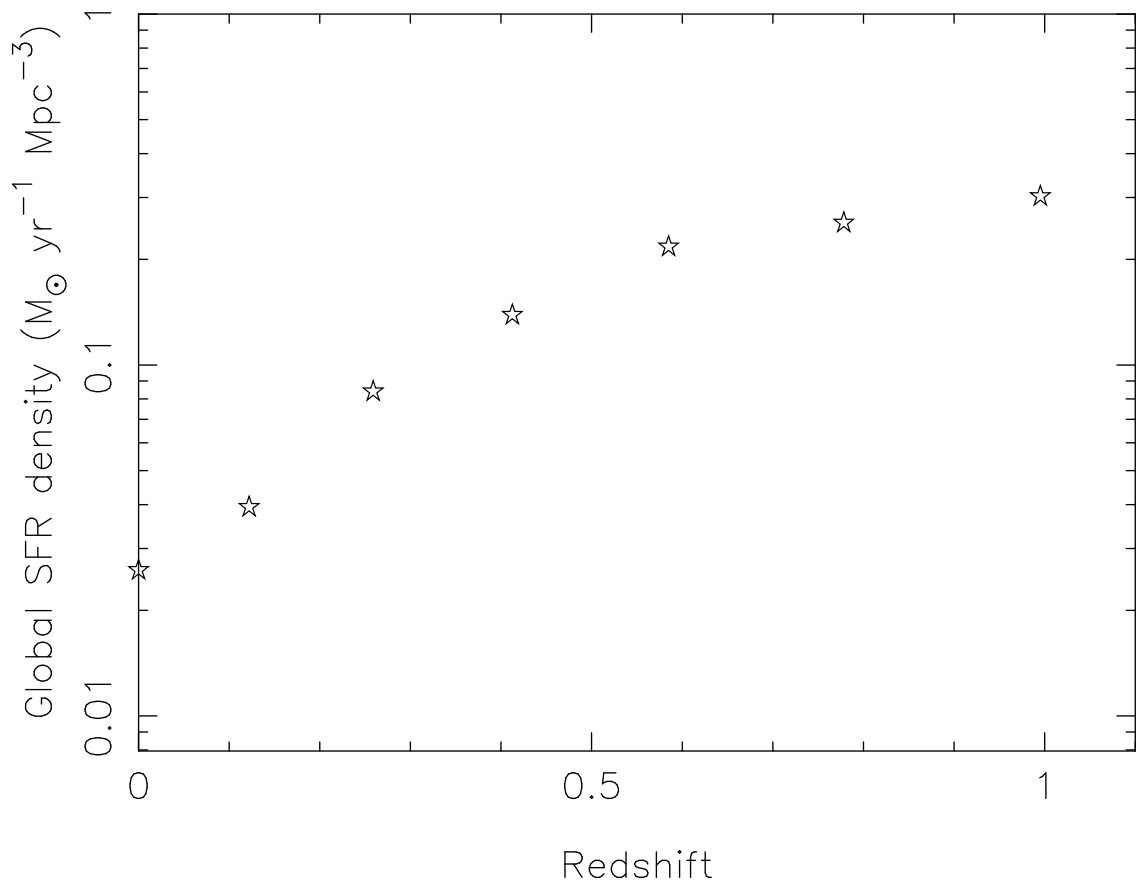
  \figcaption[Fig1.epsi]{Variation of the global star formation rate
    density with redshift, derived from the 1.4~GHz luminosity function of
    Condon (1984b).}
\end{minipage}
\end{center}

\newpage
\begin{center}
\begin{minipage}{15cm}
  \psfig{figure=Fig2.ps,angle=-90,width=15cm}
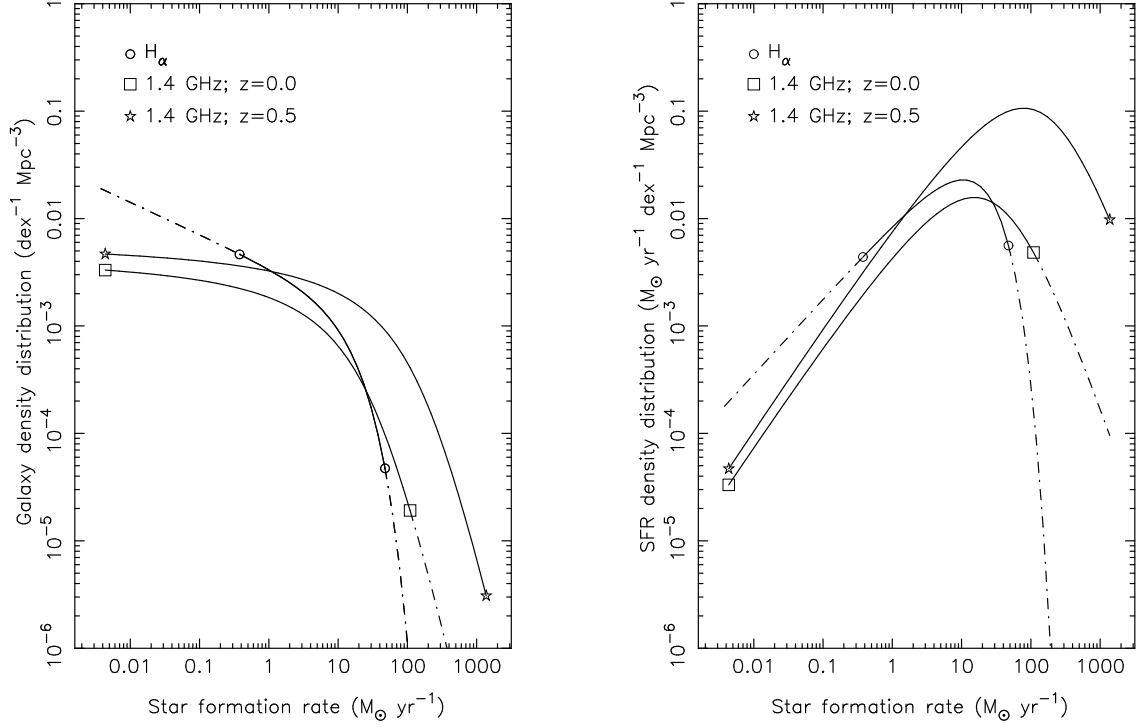
  \figcaption[Fig2.epsi]{Luminosity distributions (left panel) and star
    formation rate densities (right panel). Symbols denote the indicator
    used, and the redshift of the determination. The dash-dot curve depicts
    extrapolations of the local \Halpha~and 1.4 GHz distributions beyond
    the brightest and faintest detected objects. The abscissae, labelled
    ``star formation rate'', represent a common scale for 1.4~GHz and
    \Halpha~luminosities, corresponding to L$_{1.4}/2.35 \times 10^{22}$ W
    Hz$^{-1}$ and L$_{H\alpha}/9.4 \times 10^{33}$ W.}
\end{minipage}
\end{center}

\end{document}